\documentclass[11pt]{article}
\pdfoutput=1
\usepackage{amsmath}
\usepackage{amsthm}
\usepackage{authblk}
\usepackage{graphicx}
\usepackage{epstopdf}
\usepackage{subfig}
\usepackage{dcolumn}
\usepackage{hyperref}
\usepackage{threeparttable}
\usepackage{xcolor}
\author{Burak Ozdemir}
\affil{UNAM - National Nanotechnology Research Center and Institute of Materials Science and Nanotechnology, Bilkent University, Ankara 06800, Turkey}

\date{\today}
\title{Black Phosphorus and Phosphorene/Graphene Heterostructure as Alkali metal (Li, Na, and K) Ion Battery}
\begin{document}

\maketitle
\begin{abstract}
Black phosphorous is a layered material having a high capacity of 2596 mAhg$^{-1}$ as a battery electrode, however it suffers from cracking due to high volume expansion during lithiation. These cracks causes loss of electrical contact in the whole material, therefore capacity fades after further cycles of charging and discharging. One needs a support material which would not crack with lithiation of phosphorous in order to keep the electrical contact of the material. Here, we considered phosphorene sandwiched between graphene layers. By using density functional theory, we calculated voltages of lithiation, sodiation, and potasiation of black phosphorous and phosphorene-graphene heterostructure which compares well with the experimental results. We found low voltages for both black phosphorous and phosphorene-graphene heterostructure therefore these materials can be used as an anode electrode in lithium-ion, sodium-ion, and potassium-ion batteries.
\end{abstract}

\section{Introduction}

Electrochemical insertion of Li into black phosphorous has been experimentally studied. Sun et al. synthesized black phosphorous from red phosphorous which exhibited a high initial discharge and charge capacities of 2649 and 1425 mAhg$^{-1}$ being about 3.8 times of higher than the theoretical capacity of graphite (372 mAhg$^{-1}$).\cite{doi:10.1021/jp302265n} In that study, a stable capacity of 703 mAhg$^{-1}$ after 60 cycles is reported and it has been argued that formation of the final Li$_3$P structure causes cracking and crumbling as a reason for the significant capacity loss. However it has been demonstrated that a composite material consisting of black phosphorous and carbon has been shown to have good cyclability with 600 mAhg$^{-1}$ of capacity up to 100 cycles where final structure is LiP when voltage range is limited.\cite{ADMA:ADMA200602592} Recently, a composite of nanoparticle black phosphorous and graphite has been produced which demonstrated a high capacity of 2786 mAhg$^{-1}$ with 80\% of capacity retention up to 100 cycles for lithiation. In this composite material graphite layers are connected to black phosphorous layers. Here, graphite preserves the electrical contact between black phosphorous particles therefore keeping them as an active material for lithiation after higher cycles.\cite{doi:10.1021/nl501617j}

Black phosphorous has been studied with density functional theory (DFT) and it has been found that semi-emprical dispersion correction to PBE in the exchange-correlation functional (DFT-D) significantly improves the out-of-plane lattice constant and associated elastic constant $C_{33}$ over the PBE exchange-correlation functional.\cite{PhysRevB.86.035105}

Single layer of black phosphorous named phosphorene is also studied with DFT as a potential anode for Li and Na-ion batteries. NaP is found to be stable with respect to phosphorene and Na metal as predicted by DFT.\cite{C5CP01502B}

Heterostructure of phosphorene sandwiched between graphene layers has been recently studied with density functional theory (DFT). It has been reported that interaction energy of phosphorene and graphene is in the van der Waals range and electronic structures of phosphorene and graphene are preserved upon their contact.\cite{PhysRevLett.114.066803} Sun et al. showed experimentally that phosphorene sandwiched between graphene layers works as a Na-ion battery electrode and has a high reversible capacity of 2440 mAhg$^{-1}$.

In this work, we studied Li, Na, and K insertion into black phosphorous and graphene-phosphorense heterostructure by using density functional theory (DFT). We calculated voltages for both structures and compared to experiment. The reason we considered graphene-phosphorene heterostructure is that when black phosphorous is lithiated it expands and starts to crack which results in loss of electrical contact, therefore graphene will keep the electrical contact of the all material when phosphoerene sandwiched between graphene layers.

\section{Method}

\begin{figure}[t]
\centering
\includegraphics[angle=0, scale=0.6]{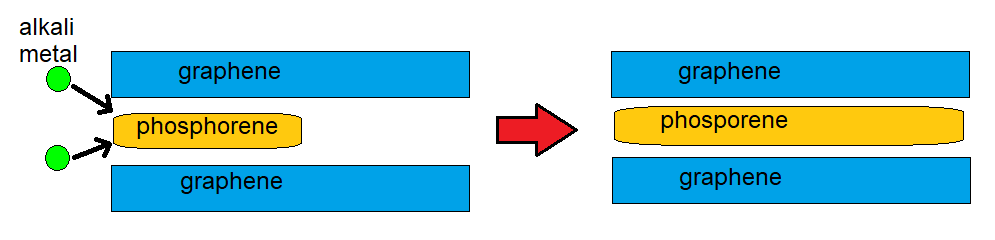}
\caption{Illustration of phosphorene expansion in between graphene layers}
\label{fexpans}
\end{figure}

Plane-wave basis-set self-consistent density functional theory calculations are carried out using Quantum-Espresso software suit. \cite{paolo} Van der Waals exchange-correlation functional vdW-DF2-C09 is employed. \cite{dion, thon, roman, klee, cooper, sabatini2012structural} Ultrasoft PBE pseudopotentials are used for phoshorous and carbon, projector augmented wave pseudopotentials are used for Li, Na. and K to replace core electrons. Monkhorst-Pack scheme \cite{monk} is used for Brillioun zone sampling. Forces are optimized using BFGS quasi-newton algorithm with a force threshold of $10^{-3}$ Ry/Bohr. For black phosphorous, the k-point grid was set to $9\times3\times7$. Kinetic energy cut-off for the wavefunctions and the charge density are set to 80 Ry and 800 Ry. Marzari-Vanderbilt smearing with 0.02 eV of smearing width is applied. Voltage of alkali metal intercalation of black phosphorous is calculated with the equation\cite{PhysRevB.82.125416};
\begin{equation}
    V(x)=-\frac{M_xP-M_{x_0}P-M(x-x_0)}{x-x_0},
\end{equation}
where M$_x$P and M$_{x_0}$P are the energies of alkali metal intercalated phosphorous, M is the energy of bulk alkali metal. However, in the case of alkali metal intercalation of graphene-phosphorene heterostructure, phosphorene expands in between graphene layers with alkali metal insertion, therefore phosphorous to carbon atom ratio changes. This is illustrated in Fig. \ref{fexpans}. In the material with less number of alkali metal atoms, phosphorene covers less area of graphene therefore there is empty region with only layered graphene. This region, which is simply graphite, needs to be cancelled in the voltage equation. Therefore a modification of the voltage equation is in order. The modified voltage equation is
\begin{equation}
    V(x)=-\frac{M_xPC_y-M_{x_0}PC_{y_0}-C(y-y_0)-M(x-x_0)}{x-x_0},
\end{equation}
where C is the energy of graphite.

\section{Results \& Discussion}

\begin{figure}[t]
\centering
\includegraphics[angle=0, scale=0.3]{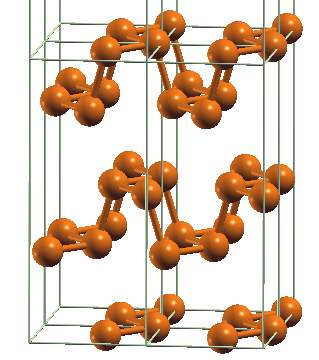}
\caption{Crystal structure of black phosphorous}
\label{fblackp}
\end{figure}

\begin{figure}[t]
\centering
\subfloat[0.285 eV (Li), 0.453 eV (Na), 0.652 eV (K)]{\label{fblackpal1}\includegraphics[angle=0, scale=0.2]{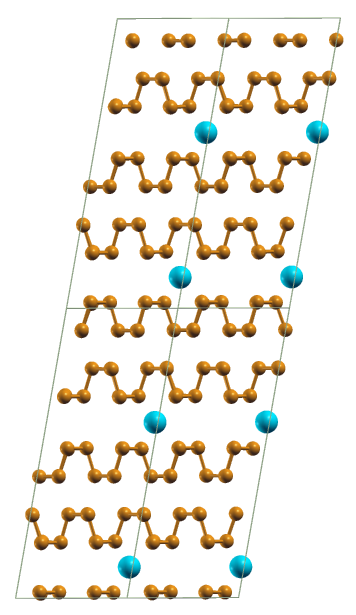}\includegraphics[angle=0, scale=0.2]{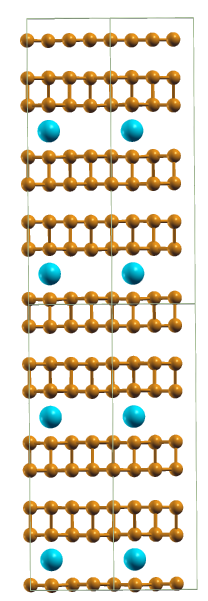}}
\subfloat[0.297 eV (Li), 0.462 eV (Na), 0.680 eV (K)]{\label{fblackpal2}\includegraphics[angle=0, scale=0.2]{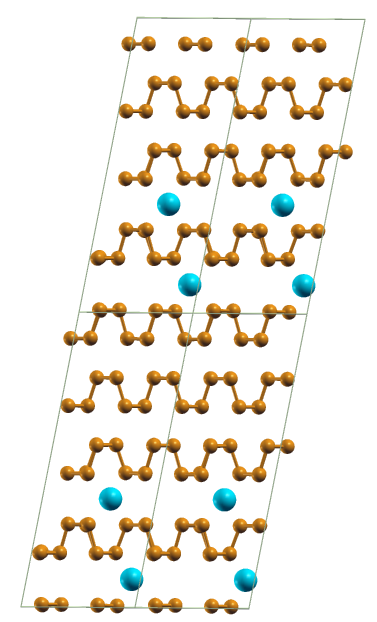}\includegraphics[angle=0, scale=0.2]{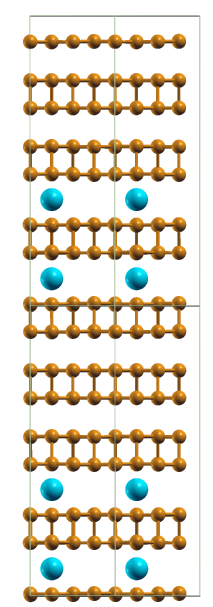}}
\subfloat[Ref.]{\label{fblackpal3}\includegraphics[angle=0, scale=0.2]{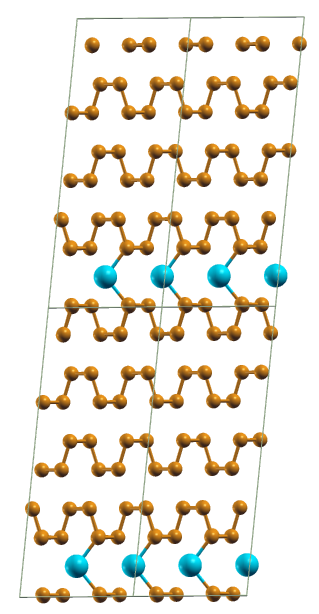}\includegraphics[angle=0, scale=0.2]{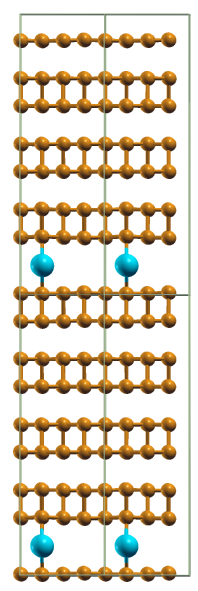}}
\subfloat[0.077 eV (Li), 0.154 eV (Na), 0.321 eV (K)]{\label{fblackpal4}\includegraphics[angle=0, scale=0.2]{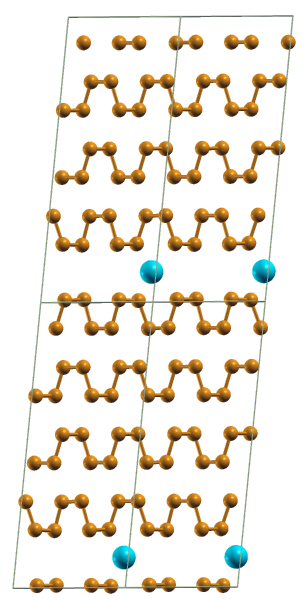}\includegraphics[angle=0, scale=0.2]{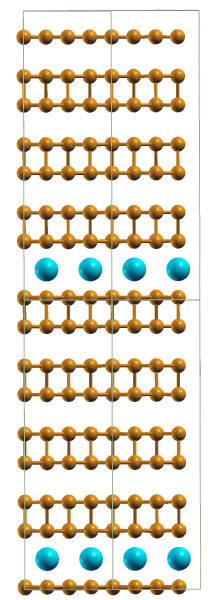}}\\
\subfloat[0.022 eV (Li), 0.045 eV (Na), -0.067 eV (K)]{\label{fblackpal5}\includegraphics[angle=0, scale=0.19]{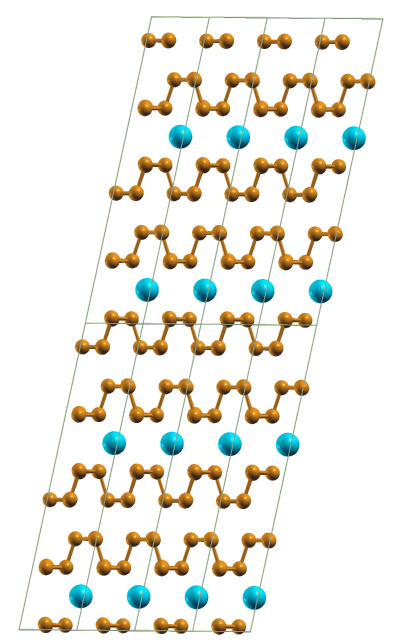}\includegraphics[angle=0, scale=0.19]{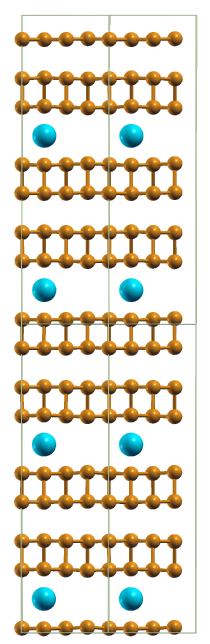}}
\subfloat[Ref.]{\label{fblackpal6}\includegraphics[angle=0, scale=0.2]{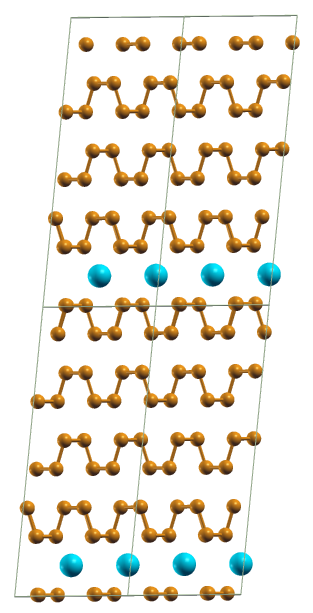}\includegraphics[angle=0, scale=0.2]{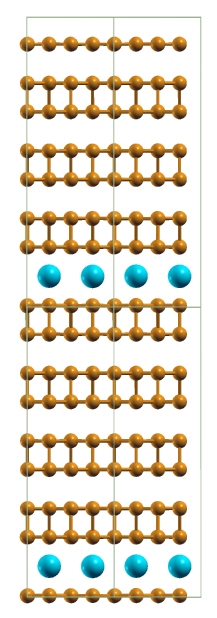}}
\caption{Crystal structures of Li, Na, and K inserted black phosphorous}
\label{fblackpal}
\end{figure}

\begin{figure}[t]
\centering
\subfloat[LiP]{\label{fig.str.2}\includegraphics[angle=0, scale=0.25]{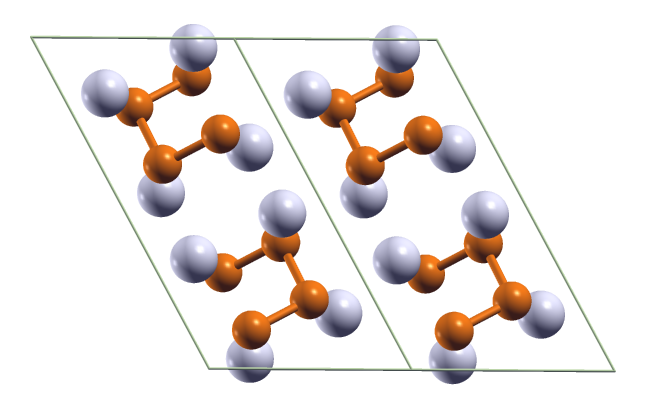}}
\subfloat[LiP]{\label{fig.str.3}\includegraphics[angle=0, scale=0.20]{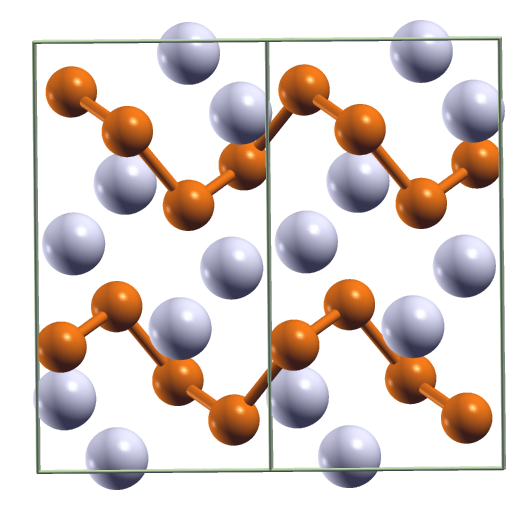}}
\subfloat[Na(K)P]{\label{fig.str.2}\includegraphics[angle=0, scale=0.25]{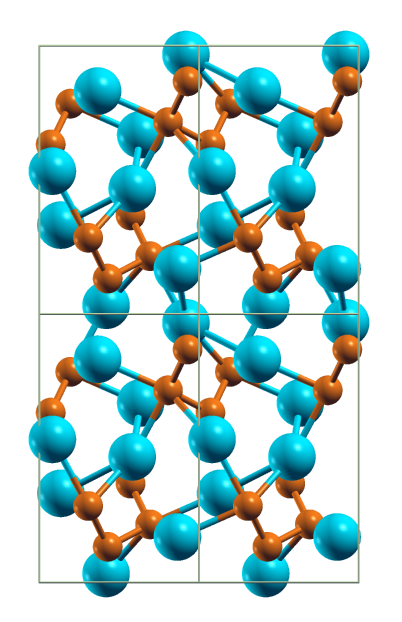}}
\subfloat[M$_3$P]{\label{fig.str.4}\includegraphics[angle=0, scale=0.20]{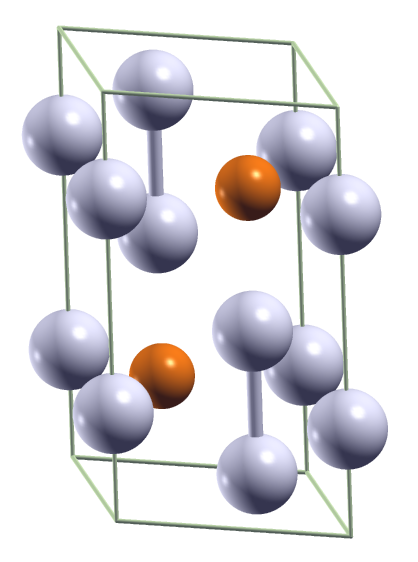}}\\
\caption{Relaxed crystal structures of (a) and (b) LiP, (c) Na(K)P, (d)  M$_3$P (M: Li, Na, K)}
\label{flip}
\end{figure}

\begin{figure}[t]
\centering
\includegraphics[angle=0, scale=1]{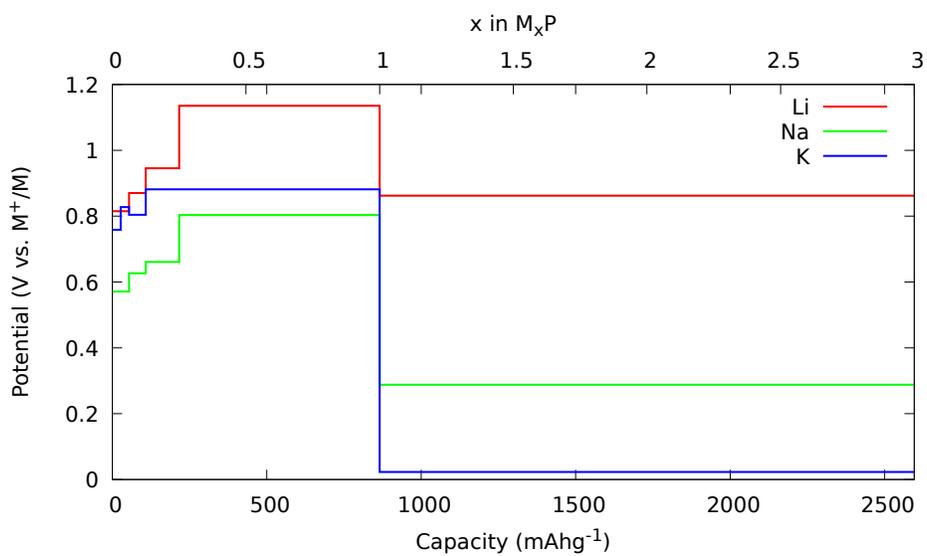}
\caption{Voltage profile of lithation, sodiation, and potasiation of black phosphorous.}
\label{fvolt}
\end{figure}

We begin our discussion with intercalation of black phosphorous. Crystal structure of black phosphorous is shown in Fig. \ref{fblackp}. Calculated lattice parameters are given in Table \ref{tlatcon} which are close to the experimental results. Several crystal structures of alkali metal intercalated black phosphorous are tried. The structures are shown in Fig. \ref{fblackpal}. Three different stages (stage-2, stage-3, and stage-4) are considered for MP$_{32}$ (M: Li, Na, K) as shown in Fig. \ref{fblackpal1}, \ref{fblackpal2}, \ref{fblackpal3}, \ref{fblackpal4}. Two different structures are considered for stage-4 as shown is Fig. \ref{fblackpal3} and \ref{fblackpal4}. Here we checked whether intercalation goes through stages or all the galleries are intercalated at once. Energy differences with respect to the reference (ref.) structure are given in Fig. \ref{fblackpal}. Energetically most stable structure is the stage-4 reference structure (Fig. \ref{fblackpal3}). Stage-4 is energetically more favorable than stage-3 and stage-2 therefore we understand that a given gallery is being intercalated until it is full of alkali metals and than another gallery is intercalated which means staging occurs in black phosphorous. Next, we considered the stoichiometry MP$_{16}$. We tried two different structures stage-4 and stage-2 for this stoichiometry. For Li and Na, stage-4 is more favorable, however for K, stage-2 is more favorable in this stoichiometry. Next, we calculated the energies of MP and M$_3$P structures which are obtained from crystallography open database (http://www.crystallography.net/cod/) and shown in Fig. \ref{flip}. Structures of NaP and KP are somewhat different than LiP. Calculated voltages of Li, Na, and K insertion into black phosphorous is given in Fig. \ref{fvolt}. First, we see increasing voltages by steps up to Li(Na)P$_4$ and KP$_8$ where we applied staging (first stage-4 then stage-2 then stage-1). Increasing voltage in this capacity range does not agree with the experimental result of Sun et al.\cite{doi:10.1021/jp302265n} where voltage of Li insertion starts from high values and decreases. We suppose that this could be due to the fact that the structures we considered here are not the energetically most favored structures and there exist another structure which is energetically more favored. Here, we performed molecular dynamics calculation and found that the structure changes and becomes amorph, however the structure found is energetically less favored therefore we did not considered in the calculation of voltage. Voltages between LiP$_4$ and Li$_3$P is around 1 V which is comparable to the experimental results.\cite{doi:10.1021/jp302265n, ADMA:ADMA200602592}. Order of voltages between MP and M$_3$P is Li$>$Na$>$K, therefore K has the lowest voltage which makes it the most suitable anode since when used with a cathode it would give the highest voltage difference therefore the highest energy density.

\begin{figure}[t]
\centering
\subfloat[P/C]{\label{fig.str.5}\includegraphics[angle=0, scale=0.2]{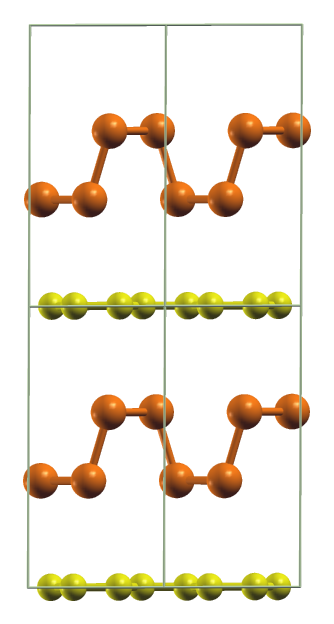}}
\subfloat[P/C]{\label{fig.str.5}\includegraphics[angle=0, scale=0.2]{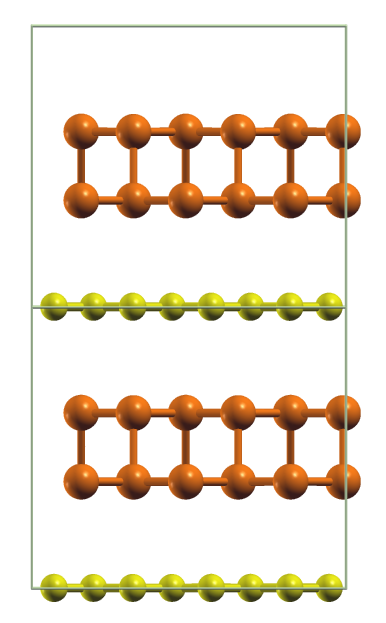}}
\subfloat[P/C]{\label{fig.str.5}\includegraphics[angle=0, scale=0.2]{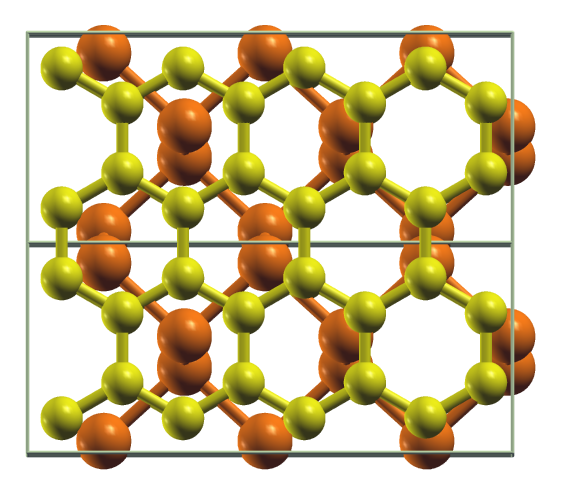}}\\
\subfloat[A$_{0.5}$P/C]{\label{fig.str.5}\includegraphics[angle=0, scale=0.2]{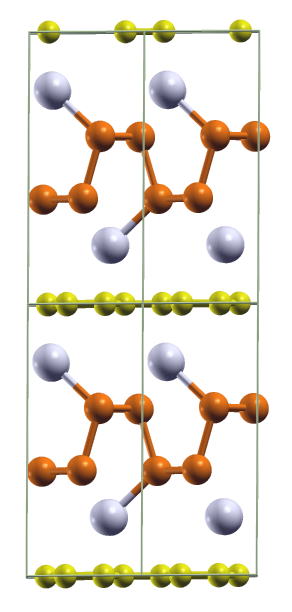}}
\subfloat[A$_{0.5}$P/C]{\label{fig.str.5}\includegraphics[angle=0, scale=0.2]{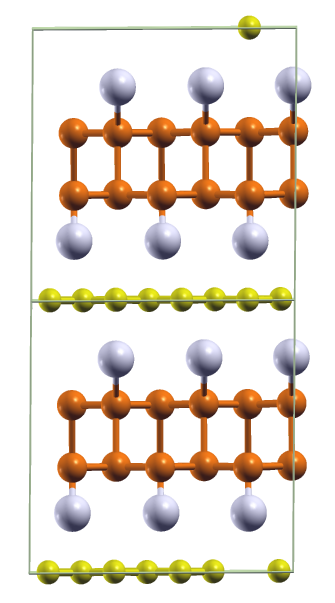}}\\
\subfloat[LiP/C]{\label{fig.str.5}\includegraphics[angle=0, scale=0.28]{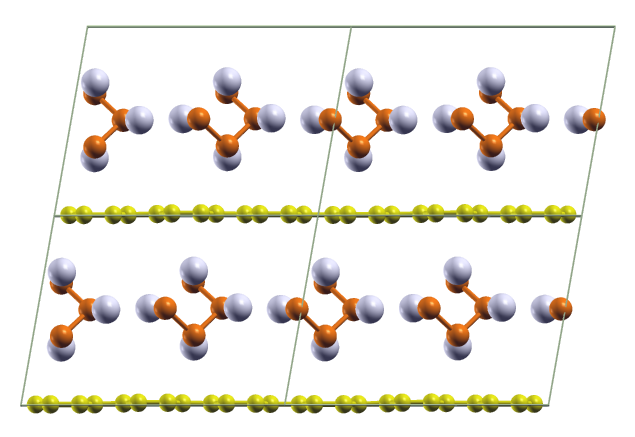}}
\subfloat[Na(K)P/C]{\label{fig.str.5}\includegraphics[angle=0, scale=0.25]{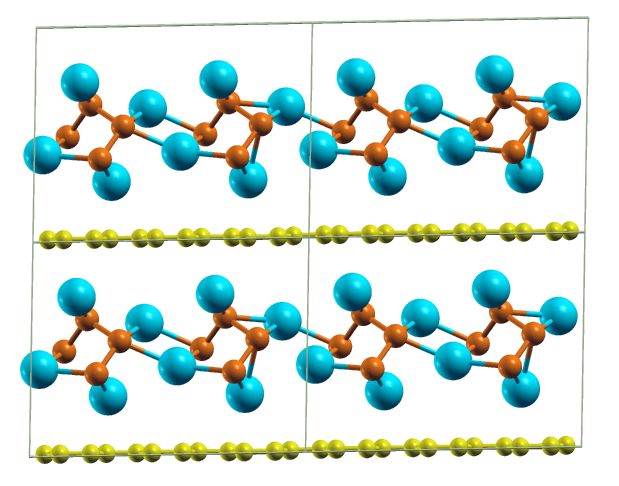}}\\
\subfloat[A$_3$P/C]{\label{fig.str.5}\includegraphics[angle=0, scale=0.2]{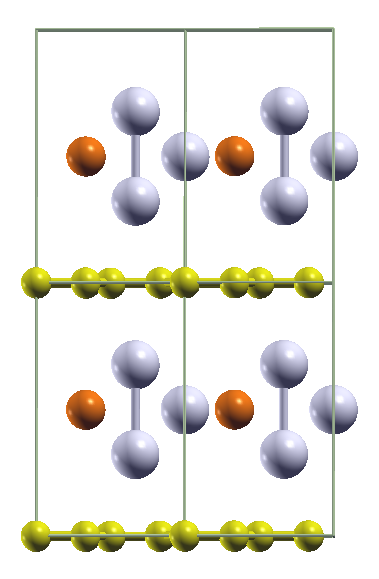}}
\subfloat[A$_3$P/C]{\label{fig.str.5}\includegraphics[angle=0, scale=0.2]{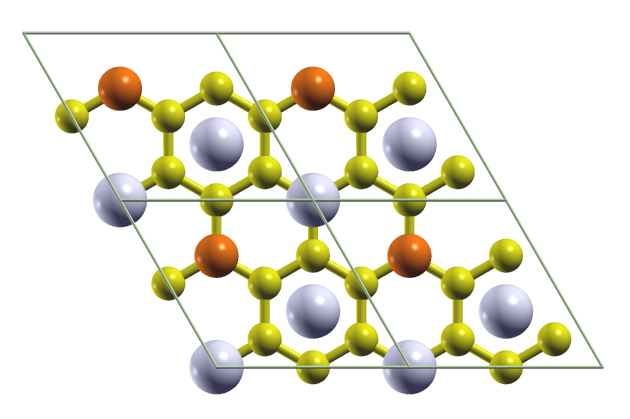}}
\subfloat[A$_3$P/C]{\label{fig.str.5}\includegraphics[angle=0, scale=0.2]{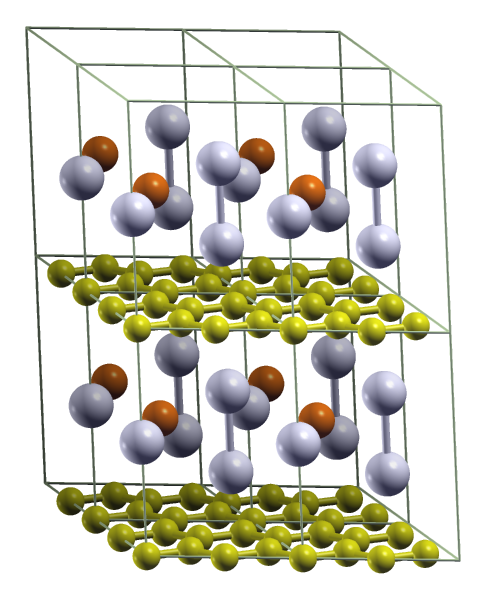}}
\caption{(a), (b), and (c) Crystal structure of graphene/phosphorene (P/C) heterostructure from three different viewing angles, (d) and (e) crystal structure of A$_{0.5}$P/C (A: Li, Na, K) from two different viewing angles, (f) crystal structure of LiP/C, (g) crystal structure of Na(K)P/C, (h), (i), and (j) are the crystal structure of M$_3$P/C from three different viewing angles}
\label{fBPC}
\end{figure}

\begin{figure}[t]
\centering
\includegraphics[angle=0, scale=1]{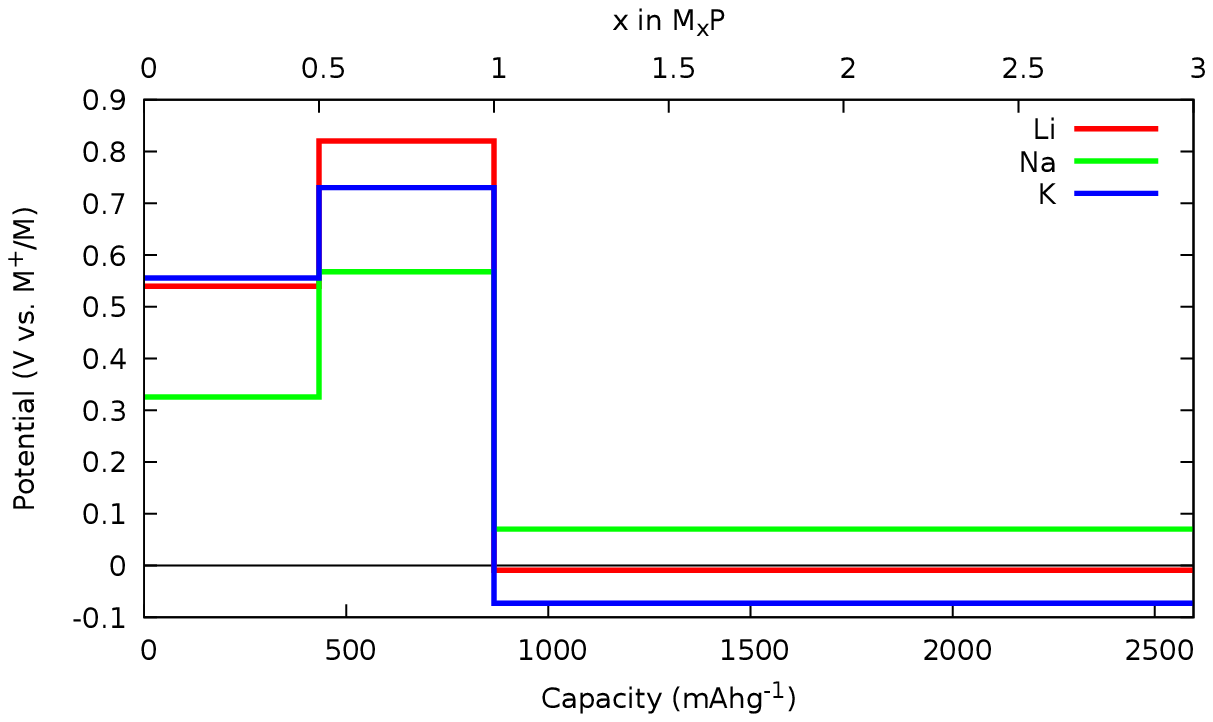}
\caption{Voltage profile of lithation, sodiation, and potasiation of phosphorene-graphene heterostructure.}
\label{fvolt1}
\end{figure}

Next, we studied Li, Na, and K insertion into graphene-phosphorene heterostructure. We considered three stoichiometries; M$_{0.5}$P/C, MP/C, M$_3$P/C. Including the graphene-phosphorene heterostructure (P/C), four structures are shown in Fig. \ref{fBPC}. For MP/C and M$_3$P/C, MP and M$_3$P structures are cut from the bulk structures and placed in between graphene layers. Here, there is a lattice mismatch between MP and graphene cell, and between M$_3$P and graphene cell. The order in amount of lattice mismatch is Li$>$Na$>$K for MP/C and K$>$Li$>$Na for M$_3$P/C. In case of MP/C, for Li, the lattice mismatch is somewhat larger than Na and K for which the mismatch is low, therefore a decrease in calculated voltage for Li is expected than its true value. And in case of M$_3$P, for K, the lattice mismatch is largest therefore the calculated voltage is expected to be lower than its true value in this capacity range. Calculated voltages are shown in Fig. \ref{fvolt1}. Results for Na insertion into phosphorene-graphene heterostructure agrees well with the experimental result of Sun et al.\cite{sun2015phosphorene} except up to the M$_{0.5}$P where voltage is increasing whereas in the experiment it starts from high values and decreases. Here we understand that there should exist another structure which is energetically more favorable. One can perform molecular dynamics calculation to find the energetically favored structure however we did not perform molecular dynamics here. Same conclusion probably applies to Li and K as well for which voltage increases up to M$_{0.5}$P. For Li and K, voltage between MP and M$_3$P is small negative value, however as we mentioned lowering of energy is expected due to lattice mismatch, therefore here the voltage is expected to be higher. Voltages are low therefore phosphorene-graphene heterostructure can be used as anode.

\begin{table}[!]
\small
\centering
\caption{Lattice constants $a$, $b$, $c$ (in \AA) of black phosphorous}
\begin{threeparttable}
\begin{tabular}{c c c c}
\hline
 & $a$ & $b$ & $c$\\
\hline\hline
vdW-DF2-C09 & 3.33 & 10.31 & 4.25\\
\hline
PBE\cite{PhysRevB.86.035105}&3.28&11.22& 4.54\\
DFT-D\cite{PhysRevB.86.035105}&3.30&10.43&4.40\\
Exp.\cite{:/content/aip/journal/jcp/71/4/10.1063/1.438523}& 3.3133 & 10.473 & 4.374\\
\hline
\end{tabular}
\end{threeparttable}
\label{tlatcon}
\end{table}

\section{Conclusion}
We performed density functional theory calculations including van der Waals interaction to study lithiation, sodiation, and potasiation of black phosphorous and phosphorene-graphene heterostructure. Lithiation of black phosphorous is studied experimentally by Sun et al.\cite{doi:10.1021/jp302265n} and Park et al.\cite{ADMA:ADMA200602592} and they found that lithiation causes significant volume expansion which causes cracks in the material therefore material looses its electrical contact after further cycles of charging and discharging. Therefore one needs a support material which does not crack with lithiation and keeps the electrical contact of the whole material. In order to solve this problem we considered phosphorene sandwiched between graphene layers which should maintain the electrical contact of the whole material after many cycles. By using DFT, we calculated the voltages of lithiation, sodiation, and potasiation of black phosphorous and phosphorene-graphene heterostructure which has not been reported before theoretically. Calculation results for lithiation of black phosphorous and sodiation of phosphorene-graphene heterostructure are in good agreement with experimental results. Also, we have showed for the first time that K can be inserted into black phosphorous and phosphorene-graphene heterostructure. Considering the high natural abundance of Na and K, and increasing demand on electrochemical batteries this is an important result. We found low voltages therefore these materials can be used as anode material.

\clearpage
\bibliographystyle{ieeetr}
\bibliography{myref}
\end{document}